\documentclass[12pt]{article}
\usepackage[left=2cm,right=2cm]{geometry}
\usepackage{graphicx}

\title{X(2175) as a resonant state of the $\phi K \bar{K}$ system}
\author{A. Mart\'inez Torres\textrm{$^1$}, K. P. Khemchandani\textrm{$^1$}\footnote{Present address: Centro de F\'isica Computacional, Deptartamento de F\' isica,  Universidade  de Coimbra, P-3004-516 Coimbra, Portugal}, L. S. Geng\textrm{$^1$},  \\
M. Napsuciale\textrm{$^{1,3}$}, and E. Oset\textrm{$^1$}}
\date{}
\begin{document}
\maketitle
\begin{center}
$^1$Departamento de F\'{\i}sica Te\'orica and IFIC,
Centro Mixto Universidad de Valencia-CSIC, Institutos de
Investigaci\'on de Paterna, Aptdo. 22085, 46071 Valencia, Spain
\end{center}

\begin{center}
$^2 $Instituto de F\'{\i}sica, Universidad de Guanajuato, Lomas
del Bosque 103, Fraccionamiento Lomas del Campestre, 37150, Le\'{o}n,
Guanajuato, M\'{e}xico.
\end{center}

\begin{abstract}
We perform a Faddeev calculation for the three mesons system, $\phi K \bar{K}$,
taking the interaction between two pseudoscalar mesons and between a vector and a pseudoscalar meson 
from the chiral unitary approach.
We obtain a neat resonance peak around a total mass of 2150
MeV and an invariant mass for the $K \bar{K}$ system around 970 MeV, very close
to the $f_0(980)$ mass. The state appears in I=0 and qualifies as a $\phi f_0(980)$
resonance. We enlarge the space of states including $\phi \pi \pi$, since $\pi \pi$ and $K \bar{K}$ 
build up  the $f_0$ (980), and find moderate changes that serve to quantify theoretical uncertainties.
No state is seen in I=1. This finding provides a natural explanation
for the recent state found at BABAR and BES, the X(2175),  which decays into $\phi f_0(980)$.

\end{abstract}

\maketitle

\section{Introduction}

The discovery of the X(2175) $1^{--}$ resonance in $e^+ e^- \to \phi f_0(980)$ with
initial state radiation at BABAR \cite{Aubert:2006bu,Aubert:2007ur}, also
confirmed at BES in  $J/\Psi \to \eta \phi f_0(980)$ \cite{:2007yt}, has
stimulated research around its nontrivial nature  in terms of quark components. 
The possibility of it being a  tetraquark $s\bar{s} s \bar{s}$ is
investigated within QCD sum rules in \cite{Wang:2006ri}, and  as a 
 gluon hybrid $s \bar{s} g$ state has been discussed in 
\cite{Ding:2007pc,Close:2007ny}. A recent review on this issue can be seen in 
\cite{Zhu:2007wz}, where the basic problem of the expected large decay widths into two mesons of the states of these models, contrary to what is experimentally observed, 
is discussed. The basic data on this resonance from 
\cite{Aubert:2006bu,Aubert:2007ur} are $M_X = 2175 \pm 10$ MeV and 
$ \Gamma = 58 \pm 16 \pm 20$ MeV,
which are consistent with the numbers quoted from BES 
$M_X = 2186 \pm 10 \pm 6$ MeV and $ \Gamma = 65 \pm 25 \pm 17$ MeV. In 
Ref. \cite{Aubert:2007ur} an indication of this resonance is seen as an increase of the $K^+ K^- K^+ K^-$ cross
section around $2150$ MeV. A detailed theoretical study of the 
$e^+ e^- \to \phi f_0(980)$ reaction was done in Ref. \cite{Napsuciale:2007wp} by
means of loop diagrams involving kaons and $K^*$, using chiral amplitudes for the
$K \bar{K} \to \pi \pi$ channel which contains the $f_0(980)$ pole generated dynamically by the theory. 
The study revealed that the loop mechanisms reproduced the background but failed to
produce the peak around $ 2175$ MeV, thus reinforcing the claims for a new
resonance around this mass.

  In the present paper, we advocate a very different picture for the X(2175) 
resonance which allows for a reliable calculation and leads naturally to a very narrow 
width and no coupling to two pseudoscalar mesons.  The picture is that the  X(2175) is 
an ordinary resonant state of $\phi f_0(980)$ due to the interaction of these
components. The $f_0(980)$ resonance is dynamically generated from
the interaction of $\pi \pi$ and $K \bar{K}$ treated as coupled channels within the
chiral unitary approach of \cite{npa,kaiser,markushin}, qualifying 
as a kind of molecule with $\pi \pi$ and $K \bar{K}$ as its components, with a large coupling to 
$K \bar{K}$ and a weaker one to $\pi \pi$ [hence, the small width compared to that of the $\sigma(600)$]. 
Similar studies for the vector-pseudoscalar interaction have also been carried out using chiral dynamics 
in \cite{lutz, Roca:2005nm}, which lead to the dynamical generation of the low-lying axial vectors. We shall follow the approach of Ref. \cite{Roca:2005nm} to deal with this part of the problem and 
will use the $\phi K$ and $\phi \pi$ amplitudes obtained in that approach.

To study
the  $\phi f_0(980)$ interaction, we are thus forced to investigate the three-body system 
$\phi K \bar{K}$ considering the interaction of the three components among
themselves and keeping in mind the expected strong correlations of the $K
\bar{K}$  system to make the $f_0(980)$ resonance. For this purpose we have
solved the Faddeev equations  with coupled channels $\phi K^+ K^-$ and 
$\phi K^0 \bar{K^0}$. The picture is later complemented with the addition 
of the $\phi \pi \pi$ state  as a coupled channel. 
 The study benefits from  previous ones on the
$\pi\bar{K} N$ and $\pi\pi N$ along with their coupled channels done in \cite{MartinezTorres:2007sr, Khemchandani:2008rk}, where many $1/2^+$, strange, and nonstrange
low-lying baryon resonances of the Particle Data Group \cite{pdg}
were reproduced. This success encourages us to extend the model of Refs. \cite{MartinezTorres:2007sr, Khemchandani:2008rk} to study the three-meson system, i.e., $\phi K\bar{K}$. One of the interesting findings of Refs. \cite{MartinezTorres:2007sr, Khemchandani:2008rk} was a cancellation of the off-shell part of the amplitudes with the genuine three-body
forces that one obtains from the same chiral Lagrangians. This simplified
technically the approach, and we shall stick to this formalism also here. 

\section{Formalism}\label{form}
To study  the $\phi K \bar{K}$ system, it is required to solve  the Faddeev equations. 
The procedure followed is (1) we solve 
coupled-channel Bethe-Salpeter equations for pseudoscalar - pseudoscalar meson (PP)
interaction as done in \cite{npa}; and for pseudoscalar-vector mesons (PV) interaction
as in \cite{Roca:2005nm}; (2) then we solve the Faddeev equations for the 
three-body, i.e., vector-pseudoscalar-pseudoscalar (VPP) mesons, system using the model developed
in \cite{MartinezTorres:2007sr}. We describe the input and the formalism for this latter part
briefly in this section.

We calculate the three-body $T$ matrix as obtained in Ref. \cite{MartinezTorres:2007sr} in terms of $T_R$, i.e.,
\begin{equation}\label{fullt}
T_R = T^{\,12}_R + T^{\,13}_R + T^{\,21}_R + T^{\,23}_R + T^{\,31}_R + T^{\,32}_R,
\end{equation}
where
\begin{equation}\label{trest}
T^{\,ij}_R=t^ig^{ij}t^j+t^i\Big[G^{\,iji\,}T^{\,ji}_R+G^{\,ijk\,}T^{\,jk}_R\Big] \,\,{\rm i \neq j\neq k = 1,2,3}
\end{equation}
corresponds to the sum of all of the diagrams with the last two $t$ matrices being
$t^j$ and $t^i$. The $t^i$($t^j$) in Eq. (\ref{trest}) denotes the two particle scattering matrix where the particle $i(j)$ is a spectator. In the chiral formalism of \cite{npa,Roca:2005nm}, these $t$-matrices in $L = 0$, which we consider here, depend on the total energy in the center of mass of the interacting pair. The $T^{ij}_R$ can be related to the Faddeev partitions, $T^i$, as
\begin{equation}
T^i=t^i\delta^3(\vec{k}^\prime_i-\vec{k}_i)+T^{ij}_R+T^{ik}_R
\end{equation}
where $T^i$ sums all of the diagrams with the particle $i$ as a spectator in the last interaction and $\vec{k}_i$ ($\vec{k}^\prime_i$) is the initial (final) momentum of the $i$th particle in the global center of mass.

The propagator $g^{ij}$ in Eq. (\ref{trest}) can be expressed as 
\begin{equation}
g^{ij}=\Bigg( \prod_{r=1}^D \frac{1}{2E_r} \Bigg) \frac{1}{\sqrt{s}-E_i
(\vec{k}_i)-E_j(\vec{k}_j)-E_k(\vec{k}_i+\vec{k}_j)+i\epsilon}\nonumber
\end{equation}
where $\sqrt{s}$ is the total energy in the global CM system, $E_l=\sqrt{\vec{k}^2_l+m^2_l}$ is the energy of the particle $l$, and $D$ is the number of particles propagating between two consecutive interactions.
The model in Ref. \cite{MartinezTorres:2007sr} has been built by writing the terms including more than two $t$-matrices by replacing the ``$g^{ij}$'' propagator by a function $G^{i\,j\,k}$, thus leading to Eq. (\ref{trest}). The function $G^{i\,j\,k}$ is given by
\begin{equation}
G^{i\,j\,k}=\int\frac{d^3 k^{\prime\prime}}{(2\pi)^3}\frac{1}
{2E_l}\frac{1}{2E_m}\frac{F^{i\,j \,k}(\sqrt{s},\vec{k}^{\prime\prime}
)}{\sqrt{s_{lm}}-E_l(\vec{k}^{\prime\prime})-E_m(\vec{k}^{\prime\prime})
+i\epsilon}, \label{eq:G}
\end{equation}
where  $i \neq j$, $j \neq k$, $i\neq l \neq m$, $\sqrt{s_{lm}}$ is the invariant mass of the $(lm)$ pair, and
$F^{i\,j\,k}$ is defined as
\begin{equation}
F^{i\,j\,k} = t^{j}(\sqrt{s_{int}} (\vec{k}^{\prime\prime})) \Biggr( \frac{g^{jk}|_{off-shell}}{ g^{jk}|_{on-shell}} \Biggr) [ t^{j}(\sqrt{s_{int}} (\vec{k}_{j^\prime})) ]^{-1}.\nonumber
\end{equation}
This $G^{i\,j\,k}$ is a loop function of a propagator, in the three-body scattering diagrams, in which the dependence on the loop variable of an anterior $t$ matrix and propagator has been included in the form of an off-shell factor  $F^{ijk}$. This simplifies technically solving Eq. (\ref{fullt}) and induces regrouping of the three-body diagrams giving six Faddeev partitions [Eq. (\ref{trest})] instead of three  (see \cite{MartinezTorres:2007sr} for a more detailed discussion).

We label $\phi$ as particle $1$ and $K$ and $\bar{K}$ as particle $2$ and $3$, respectively.
The invariant mass of the $K \bar{K}$ system $\sqrt{s_{23}}$ is taken as an input to the three-body 
calculations and is varied around the mass of the $f_0$. The $K\bar{K}$ interaction $t^1$ in this region contains the pole of the $f_0$(980) \cite{npa,kaiser}.
The other invariant masses $s_{12}$ and $s_{13}$ can be then calculated in terms of the $\sqrt{s_{23}}$
and total energy \cite{MartinezTorres:2007sr}. 
Thus, there are two variables of the calculations, i.e., the total energy and the invariant mass of the 
$K \bar{K}$ system.

We shall now discuss the input, i.e., the two-body $t$ matrices for the PP and PV mesons interaction.
For the PP case, the Bethe-Salpeter equation 
\begin{equation}\label{bs}
t = V + V \tilde{G} t
\end{equation}
has been solved for five coupled channels, i.e., $K^+K^-$, $K^0 \bar{K}^0$, $\pi^+\pi^-$, 
$\pi^0 \pi^0$, and $\pi^0\eta$. The potentials $V$ are calculated from the lowest
order chiral Lagrangian and the loops $\tilde{G}$ have been calculated 
using dimensional regularization as in \cite{npa}. 
The authors of \cite{npa,kaiser,markushin} found poles in the $t$ matrices, in the isospin 0 
sector, corresponding to the $\sigma$ and the $f_0$ resonances, and also the one corresponding to the $a_0$(980) for the isospin 1 case.
It was also found that the $f_0$ resonance is dominated by 
the $K \bar{K}$ channel and the pole for the $f_0$ appears at $\sim$ 973 MeV
even when the $\pi \pi$ channel is eliminated. The matrix element 
corresponding to the $K \bar{K} \rightarrow K \bar{K}$ scattering is used as 
an input, $t^1$, to solve  Eqs.(\ref{trest}) and (\ref{fullt}). 
In the two-body problem, the $f_0$(980) pole appears below the $K\bar{K}$
threshold. It corresponds to total energies of the $K \bar{K}$ system below $2m_k$ and 
in the momentum representation to purely imaginary kaon momenta if we take
$p_K^2 = m_K^2$ (which is not the case in a bound state). To avoid using unphysical
complex momenta in the three-body system, 
we give a minimum value of about 50 MeV/c  to the kaon
momentum in the $K \bar{K}$ center of mass system. It should be mentioned that
the results are almost insensitive to this choice of the minimum momentum. For example, a change in 
this momentum by about 40$\%$ changes the position of the peak merely by $\sim$ 5 MeV.

For the VP meson interaction, Eq. (\ref{bs}) is calculated with $\phi K$, $\omega K$, $\rho K$, $K^* \eta$, and $K^* \pi$ as coupled channels. The potential for the VP meson-meson interaction has been
obtained from the lowest order chiral Lagrangian and projected in the s wave \cite{Roca:2005nm}, and then the $\phi K \rightarrow\phi K $ element of the resulting coupled-channel $t$ matrix is used as an input in Eq.  (\ref{trest})

Coming back to the three-body problem,
we take the $\phi \bar{K} (K) \rightarrow  \phi \bar{K} (K)$ $t$-matrix element as
$t^2 (t^3)$ to solve Eqs.(\ref{trest}) and (\ref{fullt}). Our interest is to check the possibility
of existence of a resonance or a bound state with isospin zero in the $\phi K \bar{K}$ system; thus the full $T_R$ matrix [Eq. (\ref{fullt})] is to be projected to total isospin 0. 
When adding the $\phi \pi \pi$ channel, we must deal with the $\pi \pi$ and $\phi \pi$ interactions which are part of the coupled-channel study of the scalar and axial vector resonances, respectively. 

\section{A discussion on possible coupled channels}\label{sec}

In the construction of the $K\bar{K}$ and $\phi K$ two-body $t$ matrices we have used the full space of coupled channels as indicated in  Sec. \ref{form}. We shall argue here that in the three-body basis we can omit some states. The $\phi K$ system couples to 
$\omega K$, $\rho K$, $K^* \pi$, and $K^* \eta$. We shall bear in mind that we are looking for  a state with $I$ = 0 and with $\sqrt{s_{23}} \simeq$ 980 MeV, as found in the experiment \cite{Aubert:2006bu,Aubert:2007ur}. When adding the  $\bar{K}$ of the three-body $\phi K \bar{K}$ system to the coupled channels of the $\phi K$, we obtain the following states: 
$\omega K \bar{K}$, $\rho K \bar{K}$, $K^* \pi \bar{K}$, and $K^* \eta \bar{K}$. If we want the subsystem of two pseudoscalar mesons to build up the $f_0$ (980), which is dynamically generated in the $K \bar{K}$ and $\pi \pi$ interaction, we must exclude the 
$K^* \pi \bar{K}$ and $K^* \eta \bar{K}$ states. The $\rho K \bar{K}$ state is also excluded because when $K \bar{K}$ couples to the $f_0$ (980) the total isospin of the state is $I$ = 1. Only the $\omega  K \bar{K}$ state is left over. We could add this channel to the $\phi K \bar{K}$, but the $\omega  K \bar{K}$ channel lies $\sim$ 400 MeV below the $X$ (2175) resonance mass and hence is not expected to have much influence in that region. In more technical words, a channel which lies far away from the energy region under investigation would only bring a small and smooth energy-independent contribution to the final amplitude because of the large off-shellness of the propagators.

Thus the introduction of the $\omega K \bar{K}$  channel can only influence mildly the results obtained with the
$\phi K \bar{K}$ system alone, and thus we neglect it in the study. Furthermore we have also seen that the $\phi K \rightarrow \omega K$ and $\omega K \rightarrow \omega K$ amplitudes are weaker than the $\phi K \rightarrow \phi K$
one. 

Even though we argue above that $\bar{K}^*\pi K$ and $\bar{K}^*\eta K$ channels should be neglected, we have also investigated the effect of including the $\bar{K}^*\pi K$ channel, as an example. This is a channel where the $\pi K$ interaction (together with the $\eta K$ channel) leads to the scalar $\kappa$ resonance, and actually there are works which hint towards  a possibility of  $\bar{K}^*\kappa$ forming a molecule with mass around 1576 MeV \cite{guo}. What we find can be summarized as follows:
\begin{itemize}
\item In the energy region of our interest, we find a small transition amplitude from $\phi K \rightarrow K^*\pi$ as compared to $\phi K \rightarrow \phi K $, indicating a small mixture of the $\phi K\bar{K}$, and $\bar{K}^*\pi K$ components.
\item Studying the $\bar{K}^*\pi K$ system alone, we find that the corresponding amplitudes are much smaller in size than those found in the $\phi K\bar{K}$ system in the energy region around 2150 MeV.
\item In the region of energies around 1600 MeV, the $\bar{K}^*\pi K$ amplitudes can be bigger than around 2150 MeV, but they are still smaller than the  $\phi K\bar{K}$ amplitude at 2150 MeV.
\end{itemize} 

From these findings we conclude that, although more detailed work needs to be done at energies around 1600 MeV to check the suggestion of \cite{guo}, the amplitude of the $\bar{K}^*\pi K$ channel in this energy region seems too weak  to support bound states. On the other hand, we can be more assertive by stating that the effect of the $\bar{K}^*\pi K$ channel around 2150 MeV is negligible.

We can now stick to having the $\phi$ as the vector meson and  $K \bar{K}$ as the main meson-meson channel. Yet,  $K \bar{K}$ and $\pi \pi$ are strongly coupled in $I$=0,  both the  $K \bar{K} \rightarrow K \bar{K}$ and $\pi \pi \rightarrow \pi \pi$ amplitudes are strong, and it is only the intricate nonlinear dynamics of coupled channels of the Bethe-Salpeter equations that produces at the end two states, the $\sigma$ that couples strongly to the $\pi \pi$ channel and the $f_0$ (980) that couple strongly to $K \bar{K}$. Hence, we find advisable to include $\phi\pi\pi$  as a
coupled channel.

\section{Results}
\begin{figure}[ht]
\begin{center}
\includegraphics[scale=0.7]{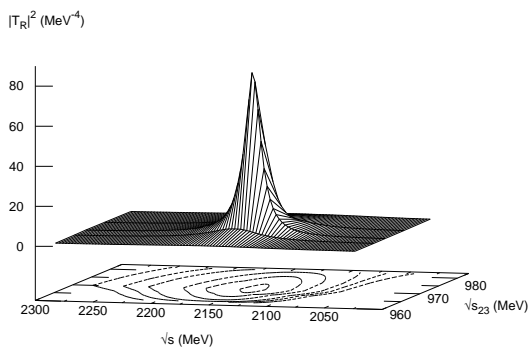}
\caption{The $\phi K \bar{K}$ squared amplitude in the isospin 0 configuration.}\label{ampsq}
\end{center}
\end{figure}

In Fig. \ref{ampsq}, we show the squared amplitude $\mid T_R \mid^2$ and its projection, as a function of the total energy ($\sqrt{s}$) and the
invariant mass of the $K\bar{K}$ system ($\sqrt{s_{23}}$), in the isospin zero configuration.
We have made the isospin projection of the amplitude of Eq. (\ref{fullt})
 using the phase convention $\mid K^- \rangle = -\mid 1/2,-1/2 \rangle$ as
\begin{equation}
\mid\phi K\bar{K};I=0,I_{K\bar{K}}=0\rangle=\frac{1}{\sqrt{2}}\Big[\mid\phi K^+ K^-\rangle +\mid\phi K^0 \bar{K}^0\rangle\Big].\nonumber
\end{equation}

A clear sharp peak of $\mid T_R\mid^2$ can be seen at 
2150 MeV, with a full width at half maximum $\sim$ 16 MeV. In order to make a meaningful comparison of this width with the experimental results, we have folded the theoretical distribution with the experimental resolution of about 10 MeV and then we find an appropriate Breit-Wigner distribution with a width $\Gamma \sim$ 27 MeV. The peak in $\mid T_R\mid^2$ appears for the $\sqrt{s_{23}}$ $\sim$ 970 MeV which is very close to the pole of the $f_0$ resonance \cite{npa}.

The total mass, the invariant mass of the  $K \bar{K}$ subsystem and the quantum numbers $I^G J^{PC} = 0^- 1^{--}$ of the resonance found here are all in agreement with those found experimentally for the X(2175) \cite{Aubert:2006bu, Aubert:2007ur}. These findings strongly suggest that this resonance can be identified with the $X(2175)$. 

Yet, our approach can go further and we can make an evaluation of the production cross section and compare it with the experimental results of \cite{Aubert:2006bu, Aubert:2007ur}. For this we make use of the theoretical evaluation of the $\phi f_0 (980)$ production in the $e^+ e^-$ reaction studied in \cite{Napsuciale:2007wp}. The authors in \cite{Napsuciale:2007wp} studied the production of the $\phi$ and $f_0 (980)$  as plane waves ($pw$) in the final state and could reproduce the background but not the peak structure around $X(2175)$ mass. Since our resonance develops from the interaction of the $\phi$ and $f_0$, the consideration of the final state interaction ($fsi$), in addition to the uncorrelated $\phi f_0$ production amplitude ($T_{pw}^{\phi f_0}$) of \cite{Napsuciale:2007wp}, could explain the experimental data in the peak region. 
We show here that this is indeed the case. We implement the $\phi f_0$ $fsi$
by multiplying $T_{pw}^{\phi f_0}$ by the factor \cite{marco, vanbeveren, roca, meissner}
\begin{equation}
 F_{fsi} = [ 1 + \tilde{G}_{\phi f_0}(s) t_{\phi f_0}(s) ],\label{fsi}
\end{equation}
where $t_{\phi f_0}$ is the scattering matrix for $\phi$ and $f_0$ and $\tilde{G}_{\phi f_0} (s)$ is the loop function of the $\phi$ and $f_0$ propagators. For $\tilde{G}_{\phi f_0}$ we use the standard formula for two mesons \cite{npa} with a cut-off ($\Lambda$) of the order of the sum of the two meson masses, as was the case in \cite{npa}, and hence $\Lambda \sim 2$ GeV here. We do not have the $t_{\phi f_0}$, but in the vicinity of the resonance it must be proportional to the three-body $T_R$ [Eq. (\ref{fullt})], implying $T_{\phi f_0} = \alpha T_R$. The proportionality coefficient  $\alpha$ is readily obtained using a relation based on unitarity, $Im \{T^{-1}_{\phi f_0}\} = - Im\{ \tilde{G}_{\phi f_0}\}$, implicit in Eq. (\ref{bs}). Assuming the $\phi f_0$ channel to be the main source of $Im \{T_R\}$, as the experimental study suggests \cite{Aubert:2006bu,Aubert:2007ur}, we have
\begin{equation}
  Im\{T^{-1}_{\phi f_0}\} = \alpha^{-1} Im\{T_R^{-1}\} = -Im\{\tilde{G}_{\phi f_0}\} = \frac{k_\phi}{8 \pi
  \sqrt{s}},\label{unitarity}
\end{equation}
which determines $\alpha$. In Eq. (\ref{unitarity}), $k_\phi$ is the $\phi$ momentum in the $\phi f_0$ center of mass system.
\begin{figure}[ht]
\begin{center}
\includegraphics[scale=0.3]{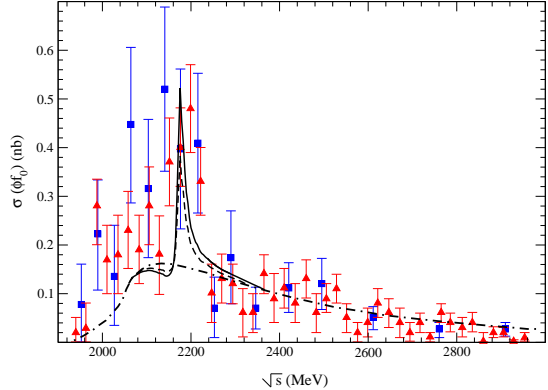}
\caption{The cross section for the $e^+ \,e^- \, \rightarrow \,\phi \, f_0$ reaction. The dashed-dotted line shows the result of the calculation of the cross section in the plane wave approximation \cite{Napsuciale:2007wp}. The dashed (solid) line shows the result of multiplying the amplitude from Ref. \cite{Napsuciale:2007wp} by the final state interaction factor [Eq. (\ref{fsi})] calculated using a cutoff of 2 (2.5) GeV for the $\tilde{G}_{\phi f_0}(s)$. The data, which corresponds to the $e^+ \,e^- \, \rightarrow \,\phi \, (\pi \pi)_{I=0}$ reaction (triangles for charged pions and boxes for neutral pions), have been taken from \cite{Aubert:2006bu, Aubert:2007ur}.}\label{crosn}
\end{center}
\end{figure}

With this information we evaluate the $e^+ \, e^- \, \rightarrow \, \phi \, f_0$ production cross section taking the results for the $\phi f_0$ production in the plane wave approximation from \cite{Napsuciale:2007wp}, and we show the results in Fig.\ref{crosn}. We can see that taking a cut-off of the order of 2-2.5 GeV for the $\tilde{G}_{\phi f_0}$, we obtain results for the production cross section which are in fair agreement with the experimental ones. In order to compare the results with the experimental cross sections in the X (2175) mass region, the energy argument of the amplitude $T_R$ has been shifted by $\sim 25$ MeV. Note, however, that the difference of $25$ MeV in the energy position (1 $\%$ of the mass) represents a remarkable agreement for a hadronic model of meson spectra.

\begin{figure}
\begin{center}
\includegraphics[scale=0.7]{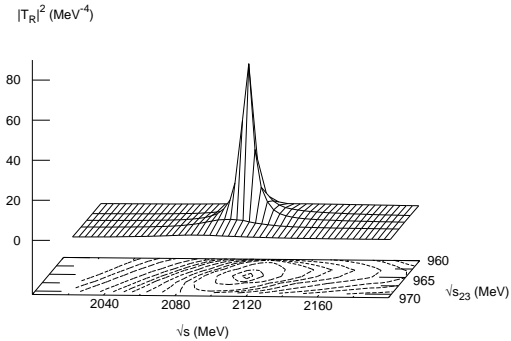}
\caption{The squared amplitude in the isospin 0 configuration including the $\phi\pi\pi$ channel.}\label{ampsqpi}
\end{center}\end{figure}

We would like now to comment on the effects of including the $\phi\pi\pi$ channel, as discussed in Sec.  \ref{sec} . We observe a similar peak as in Fig. \ref{ampsq} (see Fig. \ref{ampsqpi}); however, the position  of the peak in the total energy has been displaced by about 38 MeV downwards to an energy of 2112 MeV. At the same time, the peak shows up around $\sqrt{s_{23}} \simeq$ 965 MeV, about 15 MeV below the nominal energy of the $f_0$ (980). These differences with the nominal values of the masses of the resonances are typical of any hadronic model of resonances and, thus, the association of the resonance found to the $X$ (2175), which has the same quantum numbers as the resonance found, is the most reasonable conclusion.
In any case, the different options taken along the work, have always led to a clean peak around the same position, and the difference found could give us an idea of the theoretical uncertainties.

Finally, it should be mentioned that the $T_R$ matrix for the isospin = 1 
does not show any structure.

We have checked the sensitivity of the resonance found to the change in the cut-off ($\Lambda \sim$ 1000 MeV) used in the calculation of the input two-body $t$ matrices [Eq. (\ref{bs})], which gives the same results as dimensional regularization. There is not much freedom to change the $\Lambda$ in this case, because it has been constrained by reproducing the data on the respective two-body scattering. We thus change $\Lambda$ by $\sim$ 20 MeV for calculating Eq. (\ref{bs}), which still guarantees a reasonable agreement with the two-body cross sections, and find that it gives rise to a change in the peak position (in Fig.\ref{ampsq}) in $\sqrt{s}$ by $\sim$ 8 MeV. The cutoff is also needed to evaluate  the $G$ functions of Eq. (\ref{eq:G}), and we use the same cut-off of about 1 GeV. Since this function involves loops with three-meson propagators, it is very insensitive to the cutoff. The same change of $\sim 20$ MeV (or more) in  $\Lambda$ leads to negligible changes in the results in this case.

\section{Off-shell effects and three-body forces}

Our approach makes explicit use of the cancellation of the off-shell parts of the two-body $t$ matrices in the three-body diagrams with the genuine three-body forces, which arise from the same chiral Lagrangians
. The off-shell part of a scattering matrix is unphysical and can be changed arbitrarily by performing a unitary transformation of the fields. 

Inside the loops, the off-shell part of the chiral amplitudes, which behaves as $p^2 -m^2$ (where $p$ is the four vector of the off-shell particle) for each of the meson legs, cancels a propagator leading to a diagram with the topology of a three-body force \cite{MartinezTorres:2007sr}. It is also a peculiar feature of the chiral Lagrangians that there is a cancellation of these three-body forces with those arising from the $PPV \rightarrow PPV$ contact terms of the theory. Examples of similar cancellations are well known in chiral theories \cite{carmen,ulf}.  The detailed derivation of the cancellation of the off-shell part of the $t$ matrices and the three-body force arising from the chiral Lagrangian can be seen in the appendix \cite{Khemchandani:2008rk} for the $\pi\pi N$ interaction. In the present case, the cancellation also occurs, but it is slightly different technically. In fact, its derivation is easier than in the case studied in  \cite{Khemchandani:2008rk}, and we discuss it in the appendix  for a case in which the leading order contribution to the $V P$ transition is not null. However, in the $\phi K  \rightarrow \phi K $ case, the potential is zero. In this case, the $t$ matrix is generated by rescattering through $K^*\pi$ and $K^*\eta$ states, and the cancellation is  found in the transition potentials.

We find also instructive to see what one gets if the off-shell part of the
two-body $t$matrices is retained. Following Refs. \cite{npa, Roca:2005nm} we find, for the s wave,
\begin{eqnarray}\label{voff}
V_{K \bar{K}} (I = 0) = - \frac{1}{4f^2} (3 s_{23} - \sum\limits_i (p_i^2 - m_i^2))\\
V_{\phi K} \simeq \frac{3}{2} s_{12} - \frac{1}{2} \sum m_i^2 - \frac{1}{2} \sum\limits_i  (p_i^2 - m_i^2).
\end{eqnarray}

The $(p_i^2 - m_i^2)$ terms in Eq. (\ref{voff}) are ineffective in the loops of the two-body $t$ matrix [Eq. (\ref{bs})] \cite{npa} but will show up in the external legs of the two-body $t$ matrix used as an input in the Faddeev equations. Hence
\begin{eqnarray}\label{toff}
t_{K \bar{K}} (I = 0) = t_{on} \biggr( 1 - \frac{ \sum\limits_i (p_i^2 - m_i^2)}{3 s_{23}} \biggr) \\
t_{\phi K} = t_{on} \biggr( 1 - \frac{ \sum\limits_i  (p_i^2 - m_i^2)}{3 s_{12} - \sum\limits_i m_i^2} \biggr),
\end{eqnarray}
where $t_{on}$ denotes the corresponding on-shell $t$-matrix.
If we use these amplitudes, instead of the on-shell ones we find a very similar result to that depicted in Fig.\ref{ampsq},  with the  amplitude peaking at $\sqrt{s} = 2110$ MeV and $\sqrt{s_{23}} = 975$ MeV.
Thus, the $K\bar{K}$ still appears very correlated around the $f_0 (980)$, but the total energy has been shifted by 40 MeV. This is the result we obtain by using the off-shell $t$ matrices and neglecting the effect of the $PPV \rightarrow PPV$ contact term of the theory, which as mentioned above cancels the effect of the off-shell part of the $t$ matrix.
In other words, we could say 
that the three-body forces of the chiral Lagrangian are responsible for a shift of the resonance mass from $2110$ to $\sim$ $2150$ MeV, thus leading to a better agreement with the mass of the $X(2175)$, but, of course, the result holds for the particular choice of fields of the ordinary chiral Lagrangians.

\section{Pole in the complex plane}
One might want to see if a peak obtained in the three-body $T$ matrix corresponds to a pole in the complex plane.  The peak is so clean and close to a Breit-Wigner for a fixed $\sqrt{s_{23}}$ that it can only be reflected by a pole in the complex plane. Yet, we have looked at it in more detail through in a simplified way.  We keep the variable $\sqrt{s_{23}}$ as real, and we fix its value to the one where the peak appears and then study the $\phi K \bar{K}$ amplitude as a function of the complex $\sqrt{s}$ variable. We must move to the second Riemann sheet in the $\phi  f_0$ (980) amplitude which is accomplished by changing $k_\phi \rightarrow - k_\phi$ in the $\phi f_0$ loop function. We proceed as explained below. 

The unitarity condition allows us to write \cite{ollernsd}
\begin{equation}
T^{-1}_{\phi f_0} = V^{-1}_{\phi f_0} - \tilde{G}_{\phi f_0}
\end{equation}
with $V_{\phi f_0}$ the real potential and $\tilde{G}_{\phi f_0}$ the $\phi f_0$ loop function used in Eq. (\ref{fsi})

Going to the second Riemann sheet implies substituting $\tilde{G}_{\phi f_0}$ by $\tilde{G}^{II}_{\phi f_0}$, where $\tilde{G}^{II}_{\phi f_0}$ is obtained changing  $k_\phi$ by $-k_\phi$ in the analytical expression of $\tilde{G}_{\phi f_0}$ \cite{Roca:2005nm}
\begin{eqnarray}
&&\tilde{G}_{\phi f_0} (\sqrt{s}) = \frac{1}{16 \pi^2} \Biggr\{ a(\mu) + ln \frac{m_\phi^2}{\mu^2} + \frac{m_{f_0}^2 - m_\phi^2 + s}{2s} ln \frac{m_{f_0}^2}{m_\phi^2} + \frac{k_\phi}{\sqrt{s}} \\ \nonumber 
&&\big[ ln(s - (m_\phi^2 - m_{f_0}^2) + 2k_\phi\sqrt{s}) + ln (s + (m_\phi^2 - m_{f_0}^2) + 2k_\phi\sqrt{s}) -  \\ \nonumber
&&ln (s - (m_\phi^2 - m_{f_0}^2) - 2k_\phi\sqrt{s})
- ln (s + (m_\phi^2 - m_{f_0}^2) - 2k_\phi\sqrt{s}) - 2 \pi i \big] \Biggr\}
\end{eqnarray}
Thus we can write 
\begin{eqnarray}
 (T^{-1}_{\phi f_0})^{II}  &=&  (V^{-1}_{\phi f_0}) - (\tilde{G}_{\phi f_0})^{II}\\\nonumber
&=&  (T^{-1}_{\phi f_0})^{I} + \tilde{G}_{\phi f_0} - (\tilde{G}_{\phi f_0})^{II}\\\nonumber
&=&  (T^{-1}_{\phi f_0})^{I} - i \frac{k_\phi}{4 \pi \sqrt{s}},
\end{eqnarray}
where $I$ and $II$ indicate the first and second Riemann sheet, respectively. We can approximate $T_R$  of Eq. (\ref{trest}) by a Breit-Wigner form as
 \begin{equation}
 T_R \simeq \frac{g^2}{s - s_o + i M \Gamma (s)}
\end{equation}
from where, by means of Eq. (\ref{unitarity}), since $\alpha$ is real 
\begin{equation}
(T^{-1}_{\phi f_0})^{I} = (\alpha^{-1} T_R^{-1})^I = \biggr( \frac{k_\phi}{8 \pi \sqrt{s} \,Im \{T_R^{-1}\}} \biggr)_{\sqrt{s} = M} T_R^{-1}
\end{equation}
which leads to
\begin{equation}
(T^{-1}_{\phi f_0})^{II} = \biggr( \frac{k_\phi}{8 \pi \sqrt{s}} \frac{1}{M\Gamma} \biggr)_{\sqrt{s} = M} 
(s - s_o + i M \Gamma)  - i \frac{k_\phi}{4 \pi \sqrt{s}},
\end{equation}
which upon taking into account that
\begin{equation}
\Gamma = \frac{1}{8 \pi s} g^2 k_\phi,
\end{equation}
with $k_\phi$ being real, results in
\begin{equation}
(T^{-1}_{\phi f_0})^{II} = \frac{1}{g^2} \Bigg[ s - s_0 - i \frac{k_\phi}{8\pi} \Big( \frac{2 \sqrt{s} - M}{s} \Big) g^2 \Bigg]. 
\end{equation}
Then $(T^{-1}_{\phi f_0})^{II}$ has a pole at
\begin{equation}
 s - s_0 - i \frac{k_\phi}{8\pi} \Bigg( \frac{2 \sqrt{s} - M}{s} \Bigg) g^2 = 0. 
\end{equation}
which appears indeed very close to $Re \sqrt{s} \simeq \sqrt{s_0}$ and $Im \sqrt{s} \simeq \Gamma/2$ as we have checked numerically,
taking $s_0$ and $g^2$ from the shape of $T_R$. We also get the complex conjugate pole taking another branch of the logarithm.

\section{Summary}

In summary, the interaction of the $\phi K \bar{K}$ system studied with the Faddeev equations leads to a rearrangement of the $K \bar{K}$ subsystem as the $f_0$(980)($0^{++}$) resonance. Then, the $f_0$(980) together with the $\phi$ forms a narrow resonant $1^{--}$ state with a mass bigger than $m_\phi + 2 m_K$, which decays into
$\phi f_0$(980) and hence is most naturally associated to the recently discovered $X$(2175) resonance. The narrow width of around 27 MeV obtained here is compatible within errors with the experimental width $58 \pm 16 \pm 20$ MeV. We have also included $\phi \pi \pi$ as a coupled channel of   $\phi K \bar{K}$ and find a peak very similar to the one found with the $\phi K \bar{K}$ channel alone, except that the peak is displaced by 38 MeV down to smaller masses. We also noted that the theoretical uncertainties are of this order of magnitude.

The typical differences of our results with the experimental ones are in the range of 50 MeV for the mass and the width are roughly compatible. These are typical differences found in successful models of hadron spectroscopy. The theory also shows that there is no resonance in $\phi a_0$(980). Although a complete study of this state would require the addition of the $\phi\eta\pi$ channel, we found that the strength of the $\phi K\bar{K}$ amplitude in $I=1$ is much smaller in magnitude than that of the $\phi K\bar{K}$ in $I=0$, far away from developing a pole upon reasonable changes of the input variables. It would be most interesting to test experimentally this prediction.

\section{Acknowledgement}
This work is partly supported by
DGICYT Contract No. FIS2006-03438
and the Generalitat Valenciana. A. M. T.  acknowledges  support from the Ministerio de Educaci\'on y Ciencia. M. N. acknowledges support from CONACyT-M\'{e}xico under Project No. 50471-F.  K.P.K thanks the EU Integrated Infrastructure Initiative  Hadron Physics Project under Contract No. RII3-CT-2004-506078 and the
Funda\c{c}\~ao para a Ci\^{e}ncia e a Tecnologia of the Minist\'{e}rio da Ci\^{e}ncia,
Tecnologia e Ensino Superior of Portugal for financial support under Contract No. 
SFRH/BPD/40309/2007. 
\\
\\

\begin{center}
{\bf \Large Appendix}
\end{center}

In \cite{Khemchandani:2008rk}, a cancellation between the three-body force whose origin is in the off-shell part of the $t$ matrices and the one arising from the chiral Lagrangian was shown for  the $\pi\pi N$ system (as an example of a two-meson--one-baryon system). Here we are going to show that the same cancellation also occurs in the case of one vector and two pseudoscalar mesons. In this article, we have considered $\phi\pi\pi$ and $\phi K\bar{K}$ as the main coupled channels. However, the $\phi\pi$ and $\phi K$ ($\bar{K}$) $t$ matrices have been calculated taking the coupled channels of \cite{Roca:2005nm}, since the potentials for $\phi \pi\rightarrow\phi\pi$ and $\phi K (\bar{K})\rightarrow\phi K (\bar{K})$ are zero. This means that the $\phi\pi$ and $\phi K (\bar{K})$ interactions proceed through other coupled channels. Therefore, in order to see the mentioned cancellation for the $\phi\pi\pi$ and $\phi K\bar{K}$ channels, we must consider at least one loop for the $\phi\pi\rightarrow\phi\pi$ and $\phi K (\bar{K})\rightarrow\phi K (\bar{K}) $ interaction in which the intermediate state is one of the coupled channels of \cite{Roca:2005nm}.  The cancellations in this case have to be seen in the terms involving the transition to these intermediate states of the coupled channels. This can be done in the same way as it will be shown below, but for the sake of clarity we have taken a simple case to show the cancellation between the contribution of the off-shell part of the $t$ matrices and the contact term vector-pseudoscalar-pseudoscalar of the corresponding chiral Lagrangian. We consider the channel $\rho^+\pi^+\pi^-$, for which the $\rho^+\pi^+ (\pi^-) \rightarrow\rho^+\pi^+(\pi^-)$ transition is not zero at leading order, as an example. In order to simplify the formulation we take $\rho^+\pi^+\pi^-$ as the only channel of the system.

The interaction of a vector and any number of pseudoscalar mesons is described by the chiral Lagrangian \cite{Roca:2005nm}
\begin{equation}
\mathcal{L}=-Tr\{[V^\mu,\partial^\nu V_{\mu}]\Gamma_{\nu}\}\label{L}
\end{equation}
where
\begin{eqnarray}
V_{\mu}&=&\left(\begin{array}{ccc}\frac{1}{\sqrt{2}}\rho^0+\frac{1}{\sqrt{2}}w & \rho^+ & {K^*}^+ \\ \rho^- & -\frac{1}{\sqrt{2}}\rho^0+\frac{1}{\sqrt{2}}w & {K^*}^0 \\{K^*}^- & \bar{K}^{*\,0} & \phi\end{array}\right)_{\mu}\nonumber\\
\Gamma_{\nu}&=&\frac{1}{2}(u^\dagger\partial_{\nu}u+u\partial_{\nu}u^\dagger)\nonumber\\
u^2&=&e^{i\frac{\sqrt{2}}{f} P}\nonumber\\
P&=&\left(\begin{array}{ccc}\frac{1}{\sqrt{2}}\pi^0+\frac{1}{\sqrt{2}}\eta_{8} & \pi^+ & {K}^+ \\ \pi^- & -\frac{1}{\sqrt{2}}\pi^0+\frac{1}{\sqrt{6}}\eta_{8} & {K}^0 \\{K}^- & \bar{K}^{0} & -\frac{2}{\sqrt{6}}\eta_{8}\end{array}\right)\nonumber
\end{eqnarray}

If we expand $u$ in series up to terms containing two pseudoscalar fields $P$, we obtain
\begin{equation}
\Gamma_{\nu}=\frac{1}{4f^2}[P,\partial_{\nu} P]
\end{equation} 
and Eq. (\ref{L}) becomes
\begin{equation}
\mathcal{L}_{VP}=-\frac{1}{4f^2}Tr\{[V^\mu,\partial^\nu V_{\mu}][P,\partial_{\nu}P]\}\label{LVP}
\end{equation}

 For the case under consideration, i.e., $\rho^+\pi^+\rightarrow\rho^+\pi^+ $ and $\rho^+\pi^-\rightarrow\rho^+\pi^-$ , Eq. (\ref{LVP}) has the form
\begin{equation}
\mathcal{L}=-\frac{1}{2f^2}\Big(\partial^\mu\rho^-_{\nu}{\rho^{+}}^\nu-\rho^-_{\nu}\partial^\mu{\rho^{+}}^\nu\Big)\Big(\partial_\mu \pi^-\pi^+ -\pi^-\partial_{\mu}\pi^+\Big)
\end{equation}
\begin{figure}
\centering
\includegraphics[scale=0.7]{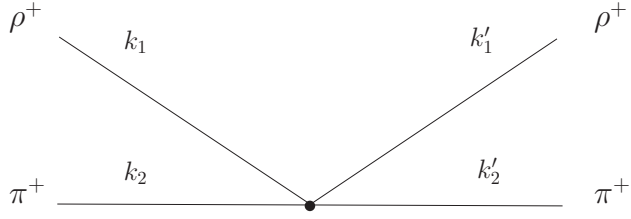}
\caption{Lowest order diagram contributing to the $\rho^+\pi^+$ interaction.}\label{fig_rhopi}
\end{figure}
 leading to (see Fig. (\ref{fig_rhopi}))
 \begin{eqnarray}
 V_{\rho^+\pi^+\rightarrow\rho^+\pi^+}&=&-\frac{1}{2f^2}(k_{1}+{k_{1}}^\prime)(k_{2}+{k_{2}}^\prime)(\epsilon\cdot\epsilon^\prime)\nonumber\\
 V_{\rho^+\pi^-\rightarrow\rho^+\pi^-}&=&-V_{\rho^+\pi^+\rightarrow\rho^+\pi^+}
  \end{eqnarray}
 From \cite{Khemchandani:2008rk}, we have 
 \begin{equation}
 V_{\pi^+\pi^-\rightarrow\pi^+\pi^-}=-\frac{1}{6f^2}\Big[3s_{\pi\pi}-\sum_{i}(k_{i}^2-m_{i}^2)\Big]
\end{equation}
where $k_{i}$ and $m_{i}$ represent the momentum and mass, respectively, of the external particles for the $\pi^+\pi^-$ interaction. 
 \begin{figure}[h!]
 \centering
 \includegraphics[scale=0.8]{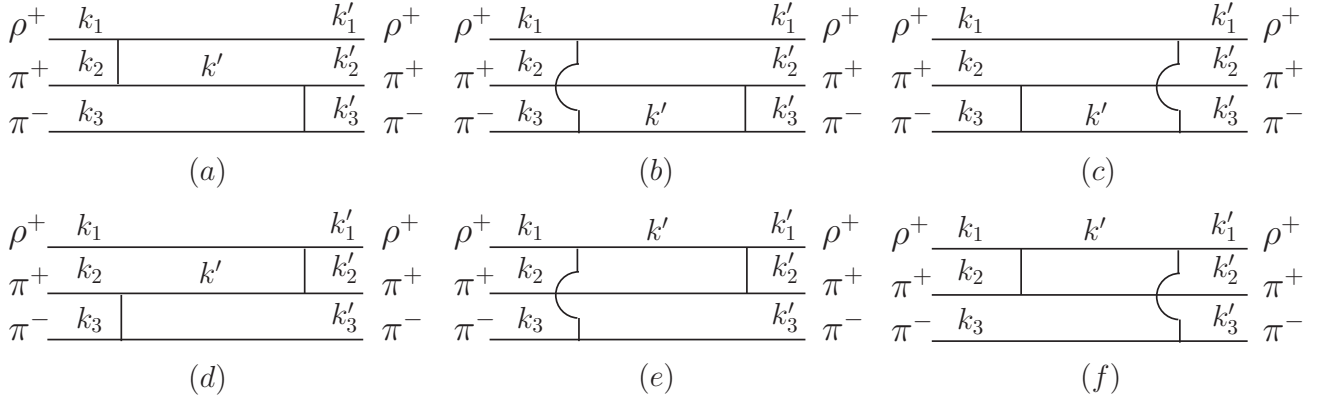}
 \caption{Diagrams in which the off-shell part of the $t$ matrices lead to a  three-body force.}\label{fig_rhopipi}
 \end{figure}

  In this way, the contribution of the first diagram in Fig. (\ref{fig_rhopipi}) is given by
 \begin{eqnarray}
 T_{a}&=&-\frac{1}{6f^2}\Big[3(k^\prime_{2}+k^\prime_{3})^2-({k^\prime}^2-m_{\pi}^2)\Big]\frac{1}{{k^\prime}^2-{m_{\pi}}^2}\Big[-\frac{1}{2f^2}(k_{1}+k^\prime_{1})(2k_2+k^\prime-k_2)\Big](\epsilon\cdot\epsilon^\prime)\nonumber\\
 &=&T^{on}_{a}+T^{off}_{a}\label{Ta},
 \end{eqnarray}
 with $T^{on}_{a}$ ($T^{off}_{a}$) being the contribution which comes from the on-shell (off-shell) part of the $t$ matrices:
 
 \begin{eqnarray}
 T^{on}_{a}&=&\frac{1}{2f^4}(k^\prime_{2}+k^\prime_{3})^2\frac{1}{(k_{1}+k_{2}-k^\prime_{1})^2-m^2_{\pi}}(k_{1}+k^\prime_{1})k_{2}(\epsilon\cdot\epsilon^\prime)\nonumber\\
 T^{off}_{a}&=&\Bigg[\frac{1}{4f^4}(k^\prime_{2}+k^\prime_{3})^2 (k_{1}+k^\prime_{1})\Bigg[\frac{k^\prime-k_{2}}{{k^\prime}^2-m^2_{\pi}}\Bigg]_{k^\prime=k_{1}+k_{2}-k^\prime_{1}}\nonumber\\&&-\frac{1}{12f^4}\frac{{k^\prime}^2-m^2_{\pi}}{{k^\prime}^2-m^2_{\pi}}(k_{1}+k^\prime_{1})(k_{2}+k^\prime)_{k^\prime=k_{1}+k_{2}-k^\prime_{1}}\Bigg](\epsilon\cdot\epsilon^\prime)
  \end{eqnarray}

 In analogy with the findings of \cite{Khemchandani:2008rk}, the contribution of the off-shell part for the different diagrams of Fig. (\ref{fig_rhopipi}), together with one of the vector-pseudoscalar-pseudoscalar contact terms of the chiral Lagrangian, is expected to vanish in the limit of equal masses for the pseudoscalars and equal masses for the vectors.  From Eq. (\ref{Ta}) and following \cite{Khemchandani:2008rk},
 \begin{equation}
 T^{off}_{a}=\Bigg[\frac{1}{4f^4}(k^\prime_{2}+k^\prime_{3})^2 (k_{1}+k^\prime_{1})\frac{\Delta k_{1}}{(\Delta k_{1})^2+2k_{2}\Delta k_{1}}-\frac{1}{12f^4}(k_{1}+k^\prime_{1})(2k_{2}+\Delta k_{1})\Bigg](\epsilon\cdot\epsilon^\prime)
 \end{equation}
 with $\Delta k_{1}=k_{1}-k^\prime_{1}$. Using that
  \begin{eqnarray}
 (k_{1}+k^\prime_{1})\Delta k_{1}&=&k_{1}^2-{k^\prime_{1}}^2=m_{1}^2-{m^\prime_{1}}^2 \end{eqnarray}
is zero in the limit of equal masses, we have
\begin{equation}
 T^{off}_{a}=-\frac{1}{6f^4}(k_{1}+k^\prime_{1})k_{2}(\epsilon\cdot\epsilon^\prime)\
\end{equation}
 By analogy, for the  rest of the diagrams in Fig. (\ref{fig_rhopipi}) we have
 \begin{eqnarray}
 T^{off}_{b}&=&\frac{1}{6f^4}(k_{1}+k^\prime_{1})k_{3}(\epsilon\cdot\epsilon^\prime)\nonumber\\
  T^{off}_{c}&=&\frac{1}{6f^4}(k_{1}+k^\prime_{1})k^\prime_{3}(\epsilon\cdot\epsilon^\prime)\nonumber\\
  T^{off}_{d}&=&-\frac{1}{6f^4}(k_{1}+k^\prime_{1})k^\prime_{2}(\epsilon\cdot\epsilon^\prime)\nonumber\\
  T^{off}_{e}&=&0\nonumber\\
  T^{off}_{f}&=&0\nonumber
 \end{eqnarray}
Adding all $T^{off}$ we get
 \begin{equation}
 \sum_{i=1}^{i=6}T^{off}_{i}=\frac{1}{6f^4}(k_{1}+k^\prime_{1})(k^\prime_{3}-k_{2}+k_{3}-k^\prime_{2})(\epsilon\cdot\epsilon^\prime)\label{Ti}
  \end{equation}
  
\begin{figure}
\centering
\includegraphics[scale=0.8]{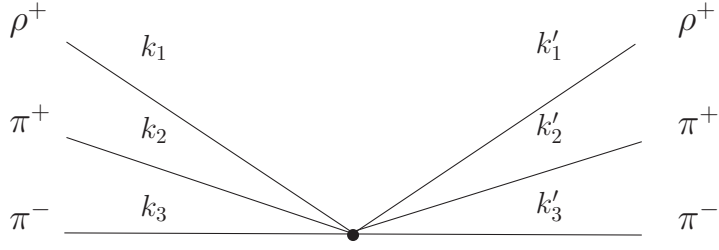}
\caption{Contact term whose origin is in the chiral Lagrangian.}\label{fig_3b}
\end{figure}

In order to evaluate the VPP contact term, we have to expand $\Gamma_{\nu}$ up to terms with four pseudoscalar fields
\begin{equation}
\Gamma_\nu=\frac{1}{32f^4}\Bigg[\frac{1}{3}\partial_\nu P P^3- P\partial_\nu P P^2+ P^2\partial_\nu P P-\frac{1}{3} P^3\partial_\nu P\Bigg]
\end{equation}
and, therefore, using Eq. (\ref{L}), the chiral Lagrangian for the VPP contact term for the $\rho^+\pi^+\pi^-$ interaction is (Fig. (\ref{fig_3b}))
 
 \begin{equation}
 \mathcal{L}_{VPP}=-\frac{1}{12f^4}(\partial^\mu\rho^-_{\nu}{\rho^+}^\nu
-\rho^-_{\nu}\partial^\mu{\rho^+}^\nu)(\pi^-\pi^-\pi^+\partial_{\mu}\pi^+ - \pi^-\partial_{\mu}\pi^-
\pi^+\pi^+)
 \end{equation}
 which implies
 \begin{equation}
 T^{3b}_{\rho^+\pi^+\pi^-}=-\frac{1}{6f^4}(k_{1}+k^\prime_{1})(k^\prime_{3}-k_{2}+k_{3}-k^\prime_{2})(\epsilon\cdot\epsilon^\prime)\label{T3}
 \end{equation}
 The sum of  Eq. (\ref{Ti}) and Eq. (\ref{T3}) results in
 \begin{equation}
 \sum_{i=1}^6T^{off}_{i}+T^{3b}_{\rho^+\pi^+\pi^-}=0
 \end{equation}

\end{document}